\documentclass[prd,twocolumn,amsfonts,showpacs]{revtex4}

\newcommand{\beq}{\begin{equation}}
\newcommand{\eeq}{\end{equation}}
\newcommand{\beqs}{\begin{eqnarray}}
\newcommand{\eeqs}{\end{eqnarray}}
\newcommand{\lsim}{\mathrel{\raisebox{-
.6ex}{$\stackrel{\textstyle<}{\sim}$}}}
\newcommand{\gsim}{\mathrel{\raisebox{-
.6ex}{$\stackrel{\textstyle>}{\sim}$}}}

\begin{document}

\title{On Condensates in Strongly Coupled Gauge Theories} 

\author{Stanley J. Brodsky$^{a,b}$}

\author{Robert Shrock$^b$}

\affiliation{(a) 
Stanford Linear Accelerator Center, Stanford University, Stanford, CA  94309}

\affiliation{(b)  
C.N. Yang Institute for Theoretical Physics, Stony Brook University, 
Stony Brook, NY 11794}

\begin{abstract}

We present a new perspective on the nature of quark and gluon condensates in
quantum chromodynamics.  We suggest that the spatial support of QCD condensates
is restricted to the interiors of hadrons, since these condensates arise due to
the interactions of confined quarks and gluons.  An analogy is drawn with order
parameters like the Cooper pair condensate and spontaneous magnetization
experimentally measured in finite samples in condensed matter physics.  Our
picture explains the results of recent studies which find no significant signal
for the vacuum gluon condensate.  We also give a general discussion of
condensates in asymptotically free vectorial and chiral gauge theories.

\end{abstract}


\maketitle

\section{Introduction} 

Hadronic condensates play an important role in quantum chromodynamics (QCD).
Two important examples are $\langle \bar q q \rangle \ \equiv \ \langle
\sum_{a=1}^{N_c} \bar q_a q^a \rangle$ and $\langle G_{\mu\nu}G^{\mu\nu}
\rangle \ \equiv \ \langle \sum_{a=1}^{N_c^2-1} G^a_{\mu \nu} G^{a \, \mu\nu}
\rangle$, where $q$ is a light quark (i.e., a quark with current-quark mass
small compared with the confinement scale), $G^a_{\mu\nu} = \partial_\mu
A^a_\nu - \partial_\nu A^a_\mu + g_s c_{abc} A_\mu^b A_\nu^c$, $a,b,c$ denote
the color indices, and $N_c=3$. (With our $(+---)$ metric, for a given $a$,
$G^a_{\mu \nu} G^{a \, \mu\nu} = 2(|{\bf B}^a|^2 - |{\bf E}^a|^2)$.)  For QCD
with $N_f$ light quarks, the $\langle \bar q q \rangle = \langle \bar q_{_L}
q_{_R} + \bar q_{_R} q_{_L} \rangle$ condensate spontaneously breaks the global
chiral symmetry ${\rm SU}(N_f)_L \times {\rm SU}(N_f)_R$ down to the diagonal,
vectorial subgroup ${\rm SU}(N_F)_{diag}$, where $N_f=2$ (or $N_f=3$ if one
includes the $s$ quark).  (Pre-QCD studies of spontaneous chiral symmetry
breaking, S$\chi$SB, include \cite{sigma,njl}.)  In an otherwise massless
theory, the $\langle G_{\mu\nu}G^{\mu\nu}\rangle$ condensate breaks dilatation
invariance.  Conventionally, these condensates are considered to be properties
of the QCD vacuum and hence to be constant throughout spacetime \cite{tmu}.

In this paper we will present a new perspective on the nature of QCD
condensates $\langle \bar q q \rangle$ and $\langle
G_{\mu\nu}G^{\mu\nu}\rangle$, particularly where they have spatial and temporal
support. We suggest that their spatial support is restricted to the interiors
of hadrons, since these condensates arise due to the interactions of quarks and
gluons which are confined within hadrons.  Chiral symmetry is thus broken in a
limited domain of size $1/ m_\pi$.  Higher-order condensates such as $\langle
(\bar q q)^2\rangle$, $\langle (\bar q q) G_{\mu\nu}G^{\mu\nu}\rangle$,
etc. are also present, and our discussion implicitly also applies to these
\cite{ggdual}.

\section{A Picture of QCD Condensates} 

We first emphasize the subtlety in characterizing the formal quantity $\langle
0 | {\cal O} | 0 \rangle$ in a canonical operator-based field theory, where
${\cal O}$ is a product of quantum field operators, by recalling that one can
render this automatically zero by normal-ordering ${\cal O}$. This subtlety is
especially delicate in a confining theory, since the vacuum state in such a
theory is not defined relative to the fields in the Lagrangian, quarks and
gluons, but to the actual physical, color-singlet, states. 

The Euclidean path-integral (vacuum-to-vacuum amplitude), $Z$, provides a
precise meaning for the expectation value $\langle {\cal O}\rangle = \lim_{J
\to 0}(\delta \ln Z/\delta J)$, where $J$ is a source for ${\cal O}$.  The path
integral for QCD, integrated over quark fields and gauge links using the
gauge-invariant lattice discretization exhibits a formal analogy with the
partition function for a statistical mechanical system.  In this
correspondence, a condensate such as $\langle \bar q q \rangle$ or $\langle
G_{\mu\nu}G^{\mu\nu}\rangle$ is analogous to an ensemble average in statistical
mechanics. It is helpful to pursue this analogy in order to
develop a physical picture of the QCD condensates.  In the context of condensed
matter physics, let us consider a phase transition which, for temperature $T <
T_c$, produces spontaneous symmetry breaking with an associated nonzero value
for some order parameter.  For example, in a superconductor, the
electron-phonon interaction produces a pairing of two electrons with opposite
spins and 3-momenta at the Fermi surface, and, for $T < T_c$, an associated
nonzero Cooper pair condensate $\langle e e \rangle_T$ \cite{fw}, where here
$\langle ... \rangle_T$ means thermal average.  Since this condensate has a
phase, the phenomenological Ginzburg-Landau (GL) free energy function $F_{GL} =
|\nabla \Phi|^2 + c_2 (\Phi^* \Phi) + c_4 (\Phi^* \Phi)^2$ uses a complex
scalar field $\Phi$ to represent it.  The formal treatment of a phase
transition in a statistical mechanical system begins with a partition function
calculated for a finite $d$-dimensional lattice, and then takes the
thermodynamic (infinite-volume) limit.  The non-analytic behavior associated
with a phase transition and the nonzero order parameter only occur in this
infinite-volume limit.  Thus, in the example of the superconductor, for $T <
T_c$, the (infinite-volume) system develops a nonzero value of the order
parameter, namely $\langle \Phi \rangle_T$, in the phenomenological
Ginzburg-Landau model, or $\langle e e \rangle_T$, in the microscopic
Bardeen-Cooper-Schrieffer theory. In the formal statistical mechanics context,
the minimization of the $|{\mathbf \nabla} \Phi|^2$ term implies that the order
parameter is a constant throughout the infinite spatial volume.

However, the infinite-volume limit is only an idealization; in reality,
superconductivity is experimentally observed to occur in finite samples of
material, such as Sn, Nb, etc., and the condensate clearly has spatial support
only in the volume of these samples.  This is evident from either of two basic
properties of a superconducting substance, namely, (i) zero-resistance flow of
electric current, and (ii) the Meissner effect, that $|{\bf B}(z)| \sim |{\bf
B}(0)|e^{-z/\lambda_L}$ for a magnetic field ${\bf B}(z)$ a distance $z$ inside
the superconducting sample, where the London penetration depth $\lambda_L$ is
given by $\lambda_L^2=m_e c^2/(4\pi n e^2)$ ($n=$ electron concentration); both
of these properties clearly hold only within the sample.  The same statement
applies to other phase transitions such as liquid-gas or ferromagnetic; again,
in the formal statistical mechanics framework, the phase transition and
associated symmetry breaking by a nonzero order parameter at low $T$ occur only
in the thermodynamic limit, but experimentally, one observes the phase
transition to occur effectively in a finite volume of matter, and the order
parameter (e.g., Cooper pair condensate or spontaneous magnetization $M$) has
support only in this finite volume, rather than the infinite volume considered
in the formal treatment. Similarly, the Goldstone modes that result from the
spontaneous breaking of a continuous symmetry (e.g., spin waves in a Heisenberg
ferromagnet) are experimentally observed in finite-volume samples.  There is,
of course, no conflict between the experimental measurements and the abstract
theorems; the key point is that these samples are large enough for the
infinite-volume limit to be a useful idealization.  Standard finite-size
scaling methods have long been used in statistical mechanics to relate real
finite-size systems and their behavior to the formal infinite-volume limit.  In
essence, one requires that the minimum dimension of the finite sample must be
large compared to the correlation length. Similar descriptions in terms of
(effective) phase transition phenomena have been applied recently to numerous
nanoscale systems, such as quantum dots \cite{dots}. 

The physics of finite-size condensed-matter systems helps to motivate our
analysis for QCD.  In our picture, the spatial support for QCD condensates is
where the color-nonsinglet particles are whose interactions give rise to them,
just as the spatial support of a magnetization $M$, say, is inside, not
outside, of a piece of iron.  This conclusion follows from the path integral
for QCD; the quark and gluon field configurations that make significant
contributions to this integral must be physical configurations, i.e., they must
be confined in hadrons.  Hence, the same should be true of the quantities
involving these fields, in particular, the quark and gluon condensates.  

The physical origin of the $\langle \bar q q \rangle$ condensate in QCD can be
argued to be due to the reversal of helicity (chirality) of a massless quark as
it moves outward from the center of a hadron and is reflected back inward 
at the boundary of a hadron, owing to confinement \cite{casher}.  This
argument implies that the condensate has support only within the spatial extent
where the quark is confined; i.e., the physical size of a hadron. Another way
to infer this is to note that in the light-front Fock-state picture of hadron
wavefunctions \cite{bpprev}, a valence quark can flip its chirality when it
interacts or interchanges with the sea quarks of multiquark Fock states, thus
providing a dynamical origin for the running quark mass.  In this description,
the $\langle \bar q q \rangle$ and $\langle G_{\mu\nu}G^{\mu\nu}\rangle$
condensates are effective quantities which originate from $q \bar q$ and gluon
contributions to the higher Fock state light-front wavefunctions of the hadron
and hence are localized within the hadron.

Let us consider a meson consisting of a light quark $q$ bound to a heavy
antiquark, such as a $B$ meson.  One can analyze the propagation of the light
$q$ in the background field of the heavy $\bar b$ quark.  Solving the
Dyson-Schwinger equation for the light quark one obtains a nonzero dynamical
mass and, via the connection mentioned above, hence a nonzero value of the
condensate $\langle \bar q q \rangle$.  But this is not a true vacuum
expectation value; instead, it is the matrix element of the operator $\bar q q$
in the background field of the $\bar b$ quark.  The change in the (dynamical)
mass of the light quark in this bound state is somewhat reminiscent of the
energy shift of an electron in the Lamb shift, in that both are consequences
the fermion being in a bound state rather than propagating freely.

Insights into the nature of spontaneous chiral symmetry breaking in QCD can be
obtained using approximate solutions of the Dyson-Schwinger equation for a
massless quark propagator; if the running coupling $\alpha_s=g_s^2/(4\pi)$
exceeds a value of order 1, this yields a nonzero dynamical (constituent) quark
mass $\Sigma$ \cite{sd}. Since in the path integral, $\Sigma$ is formally a
source for the operator $\bar qq$, one associates $\Sigma \ne 0$ with a nonzero
quark condensate (see also Refs. \cite{cjt,lnc}).  However, this application of
the Dyson-Schwinger formalism has the defect that it does not incorporate the
property of confinement, as is clear from the fact that it is an equation for a
quark, but the physical states of the theory are color-singlets, not quarks and
gluons.  This is a significant defect, since as we have noted, the
helicity-reversal of a massless quark as it is reflected inward at the outer
boundary of a hadron, due to confinement, provides a physical mechanism for
spontaneous chiral symmetry breaking in QCD.  Hence, the Dyson-Schwinger
equation for the propagator of an isolated quark cannot reliably be used to
determine where the quark condensate has spatial support; in particular, it
cannot be used to infer that it is a spacetime constant.  

Similarly, it is important to use the equations of motion for confined quarks
and gluon fields when analyzing current correlators in QCD, not free
propagators, as has often been done in traditional analyses of operator
products.  Since after a $q \bar q$ pair is created, the distance between the
quark and antiquark cannot get arbitrarily great, one cannot create a quark
condensate which has uniform extent throughout the universe.

The Anti-De Sitter/conformal field theory (AdS/CFT) correspondence between
string theory in AdS space and CFT's in physical spacetime has been used to
obtain an analytic, semi-classical model for strongly-coupled QCD which has
scale invariance and dimensional counting at short distances and color
confinement at large distances \cite{Brodsky:2008pg}.  Color confinement can be
imposed by introducing hard-wall boundary conditions at $z={1/ \Lambda_{QCD}}$,
where $z$ is the AdS fifth dimension, or by modification of the AdS metric.
This AdS/QCD model gives a good representation of the mass spectrum of
light-quark mesons and baryons as well as the hadronic wavefunctions
\cite{lightfront}.  One can also study the propagation of a scalar field $X(z)$
as a model for the dynamical running quark mass \cite{lightfront}. The AdS
solution has the form~\cite{xz} $X(z) = a_1 z+ a_2 z^3$, where $a_1$ is
proportional to the current-quark mass. The coefficient $a_2$ scales as
$\Lambda^3_{QCD}$ and is the analog of $\langle \bar q q \rangle$; however,
since the quark is a color nonsinglet, the propagation of $X(z),$ and thus the
domain of the quark condensate, is limited to the region of color confinement.
The AdS/QCD picture of condensates with spatial support restricted to hadrons
is in general agreement with results from chiral bag models \cite{chibag},
which modify the original MIT bag by coupling a pion field to the surface of
the bag in a chirally invariant manner.  Since explicit breaking of ${\rm
SU}(2)_L \times {\rm SU}(2)_R$ chiral symmetry is small, and hence $m_\pi$ is
small relative to typical hadronic mass scales like $m_\rho$ or $m_N$, these
condensates can be treated as approximately constant throughout much of the
volume of a hadron. 

Our picture of condensates with spatial support restricted to the interiors of
hadrons is consistent with the identification of pions as almost
Nambu-Goldstone bosons.  In our picture, the pions play a role analogous to the
Nambu-Goldstone modes, namely the quantized spin waves (magnons), that are
experimentally observed in a piece of a ferromagnetic substance below its Curie
temperature.  Again, strictly speaking, these spin waves result from the
spontaneous breaking of a continuous symmetry, which only occurs in an
idealized infinite-volume limit, but this limit provides a very good
approximation to a finite-volume sample.  The pions are the almost
Nambu-Goldstone bosons resulting from the spontaneous breaking of the global
${\rm SU}(2)_L \times {\rm SU}(2)_R$ chiral symmetry down to SU(2)$_{diag.}$,
and so, quite logically, the spatial support of their coordinate-space
wavefunctions is also a region where the chiral-symmetry breaking quark
condensate exists.  It is important to recall that the size of a hadron depends
not only on confinement but also on the virtual emission and reabsorption of
other hadrons, most importantly pions, since they are the lightest. Hence a
hadron can be regarded as being surrounded by a cloud of virtual pions.  By
general quantum mechanical arguments, this cloud is of size $\sim 1/m_\pi$. If
the two sources of explicit breaking of chiral ${\rm SU}(2)_L \times {\rm
SU}(2)_R$ symmetry were removed, i.e., the $m_u$ and $m_d$ current-quark masses
were taken to zero and the electroweak interactions were turned off, so that
$m_\pi=0$, then the size of a hadron, including its pion cloud, would increase
without bound (until it impinged on neighboring hadrons).  In this case, the
quark and gluon condensates would also extend throughout all of spacetime.
Thus, our picture of condensates reduces to the conventional view in the chiral
limit.  This also shows the consistency of our picture with current algebra
results such as the Gell-Mann-Oakes-Renner (GMOR) relation, $m_\pi^2 =
-f_\pi^{-2}(m_u+m_d)\langle \bar q q \rangle$ \ \cite{gmor}.  Because of
confinement, the condensate $\langle \bar q q \rangle$ has finite size $\le
1/m_\pi$.  In the chiral limit $\langle \bar q q \rangle$ becomes a constant
throughout spacetime, consistent with standard analyses. From the GMOR
relation, with the current-quark masses $m_u+m_d \simeq 12$ MeV, it follows
that $|\langle \bar q q \rangle|^{1/3} \simeq 240$ MeV. With the standard
convention that these current-quark masses are taken as positive, $\langle \bar
q q \rangle$ is negative.  As one approaches the chiral limit, in our picture,
this condensate involves a matrix element of $\bar q q$ in the nucleon, whose
size is getting very large because of its pion cloud; it thus receives
commensurately large contributions from this pion cloud around the nucleon.

We also comment on the nucleon sigma term.  Using the state normalization given
by $\langle N({\mathbf p})|N({\mathbf p}')\rangle = 2p^0 \delta^3({\mathbf p} -
{\mathbf p}')$, this term is $\sigma_{\pi N} \equiv \sigma_{\pi N}(0) =
(m_u+m_d)(2M_N)^{-1}\langle N({\mathbf p})|\bar q q | N({\mathbf
p})\rangle$. Values of $\sigma_{\pi N}$ extracted from experimental data range
from about 45 to 70 MeV \cite{schweitzer}.  A comparison of this result with
the matrix element of $\bar q q$ in the GMOR relation would be worthwhile, but,
as reviewed by Schweitzer \cite{schweitzer}, there are many theoretical and
model-dependent uncertainties.

Several studies have reported values of the (renormalization-invariant)
quantity $\langle (\alpha_s/\pi) G_{\mu\nu}G^{\mu\nu} \rangle$ by analyzing
vacuum-to-vacuum current correlators constrained by data for $e^+e^- \to$
charmonium and hadronic $\tau$ decays \cite{svz}-\cite{ggval}. In the
pioneering work on QCD sum rules \cite{svz} the authors obtained an estimate
$\simeq 0.01$ GeV$^4$. Some recent values (in GeV$^4$) include $0.006 \pm
0.012$ \cite{ggval}(a), $0.009\pm 0.007$ \cite{ggval}(b), and $-0.015 \pm
0.008$ \cite{ggval}(c).  These values show significant scatter and even
differences in sign.  In our analysis the vacuum gluon condensate vanishes; it
is confined within hadrons, rather than extending throughout all of space, as
would be true of a vacuum condensate.

In our picture, the QCD condensates should be considered as contributing to the
masses of the hadrons where they are located.  This is clear, since, e.g., a
proton subjected to a constant electric field will accelerate and, since the
condensates move with it, they comprise part of its mass. Similarly, when a
hadron decays to a non-hadronic final state, such as $\pi^0 \to \gamma\gamma$,
the condensates in this hadron contribute their energy to the final-state
photons. Thus, over long times, the dominant regions of support for these
condensates would be within nucleons, since the proton is effectively stable
(with lifetime $\tau_p >> \tau_{univ} \simeq 1.4 \times 10^{10}$ yr.), and the
neutron can be stable when bound in a nucleus.  In a process like $e^+e^- \to$
hadrons, the formation of the condensates occurs on the same time scale as
hadronization.  In accord with the Heisenberg uncertainty principle, these QCD
condensates also affect virtual processes occurring over times $t \lsim
1/\Lambda_{QCD}$.  Our suggestion implies that condensates $\langle \bar q q
\rangle$ in different hadrons may be chirally rotated with respect to each
other, somewhat analogous to disoriented chiral condensates in heavy-ion
collisions \cite{dcc}.

Lattice gauge simulations provide a powerful way to investigate properties of
QCD, including both hadron states and quark and gluon condensates \cite{lgt}.
Some early analytic and numerical lattice studies of $\langle \bar q q \rangle$
include Refs. \cite{chisb_analytic,chisb_numerical}.  Our suggestion can, in
principle, be verified by careful lattice measurements.  Note that the lattice
measurements that have inferred nonzero values of $\langle \bar q q \rangle$
and $\langle G_{\mu\nu}G^{\mu\nu}\rangle$ were performed in finite (Euclidean)
volumes, although these were usually considered as approximations to the
infinite-volume limit.

\section{Implications for General Asymptotically Free Gauge Theories} 

Having discussed QCD, we next consider, as an exercise, how our observations
apply to other asymptotically free gauge theories.  We
begin with a vectorial gauge theory with the gauge group SU($N_c$), allowing
$N_c$ to be generalized to values $N_c \ge 3$.  First, consider a theory of
this type with no fermions, so that only $\langle G_{\mu\nu}G^{\mu\nu} \rangle$
need be considered.  This condensate would then have support within the
interior of the glueballs. Second, consider a theory with $N_f=1$ massless or
light fermion transforming according to some nonsinglet representation $R$ of
SU($N_c$).  The $\langle \bar q q \rangle$ and $\langle
G_{\mu\nu}G^{\mu\nu}\rangle$ condensates in this theory would have support in
the interior of the mesons, baryons, and glueballs (or mass eigenstates formed
from glueballs and mesons).  Here, the condensate $\langle \bar q q \rangle$
does not break any non-anomalous global chiral symmetry, so there would not be
any Nambu-Goldstone boson (NGB).  In both of these theories, the sizes of the
mesons, baryons, and glueballs are $\simeq 1/\Lambda$, where $\Lambda$ is the
confinement scale.

We next consider asymptotically free chiral gauge theories (which are free of
gauge and global anomalies) with massless fermions transforming as
representations $\{R_i\}$ of the gauge group. The properties of strongly
coupled theories of this type are not as well understood as those of vectorial
gauge theories \cite{thooft}-\cite{tum}.  One possibility is that, as the
energy scale decreases from large values and the associated running coupling
$g$ increases, it eventually becomes large enough to produce a (bilinear)
fermion condensate, which thus breaks the initial gauge symmetry
\cite{tum}. This is expected to form in the most attractive channel (MAC), $R_1
\times R_2 \to R_{cond.}$, which maximizes the quantity $\Delta C_2 =
C_2(R_1)+C_2(R_2) - C_2(R_{cond.})$, where $C_2(R)$ is the quadratic Casimir
invariant.  Depending on the theory, several stages of self-breaking may occur
\cite{tum,etc}.  Let us consider an explicit model of this type, with gauge
group SU(5) and massless left-handed fermion content consisting of an
antisymmetric rank-2 tensor representation, $\psi^{ij}_L$, and a conjugate
fundamental representation, $\chi_{i,L}$.  This theory is asymptotically free
and has a formal ${\rm U}(1)_\psi \times {\rm U}(1)_\chi$ global chiral
symmetry; both U(1)'s are broken by SU(5) instantons, but the linear
combination U(1)$^\prime$ generated by $Q = Q_\psi -3Q_\chi$ is preserved.  The
MAC for condensation is $10 \times 10 \to \bar 5$, with $\Delta C_2=24/5$, and
the associated condensate is $\langle \epsilon_{ijk\ell n}\psi^{jk \ T}_L C
\psi^{\ell n}_L \rangle$, which breaks SU(5) to SU(4).  Thus, as the energy
scale decreases and the running $\alpha=g^2/(4\pi)$ grows, at a scale $\Lambda$
at which $\alpha \Delta C_2 \sim O(1)$, this condensate is expected to
form. Without loss of generality, we take $i=1$, and note
\beqs
\langle \epsilon_{1jk\ell n}\psi^{jk \ T}_L C \psi^{\ell n}_L \rangle 
& \propto & 
\langle \psi^{23 \ T}_L C \psi^{45}_L - \psi^{24 \ T}_L C \psi^{35}_L \cr\cr
& + & \psi^{25 \ T}_L C \psi^{34}_L \rangle
\label{1010condensate}
\eeqs
The nine gauge bosons in the coset SU(5)/SU(4) gain masses of order
$\Lambda$. The six components of $\psi^{ij}_L$ involved in the condensate
(\ref{1010condensate}) also gain dynamical masses of order $\Lambda$. These
components bind to form an SU(4)-singlet meson whose wavefunction is given by
the operator in (\ref{1010condensate}).  This binding involves the exchange of
the various (perturbatively massless) gauge bosons of SU(4).  The condensate
(\ref{1010condensate}) breaks the global U(1)$^\prime$, but the would-be
resultant NGB is absorbed by the gauge boson corresponding to the diagonal
generator in SU(5)/SU(4). We infer that this condensate (\ref{1010condensate})
has spatial support in the meson with the same wavefunction.  Aside from the
SU(4)-singlet $\chi_{1,L}$, the remaining massless fermion content of the SU(4)
theory is vectorial, consisting of a 4, $\psi^{1j}_L$, and a $\bar 4$,
$\chi_{j,L}$, $j=2...4$.  The formal global flavor symmetry of this effective
SU(4) theory at energy scales below $\Lambda$ is ${\rm U}(1)_L \times {\rm
U}(1)_R = {\rm U}(1)_V \times {\rm U}(1)_A$, and the U(1)$_A$ is broken by
SU(4) instantons.  This low-energy effective field theory is asymptotically
free, so that at lower energy scales, the coupling $\alpha$ that it inherits
from the SU(5) theory continues to increase, and the theory confines and
produces the condensate $\langle \psi^{1j \ T}_L C \chi_{j,L} \rangle$, which
preserves the gauged SU(4) and global U(1)$_V$.  We infer that $\langle
\psi^{1j \ T}_L C \chi_{j,L} \rangle$ and the SU(4) gluon condensate $\langle
G_{\mu\nu}G^{\mu\nu}\rangle$ have spatial support in the SU(4)-singlet baryon,
meson, and glueball states of this theory.

Although our suggestion associates condensates in a confining gauge theory $G$
with $G$-singlet hadrons, these condensates can affect properties of
$G$-singlet particles if they both couple to a common set of fields.  For
example, the $\langle \bar F F \rangle$ condensate and the corresponding
dynamical mass $\Sigma_F$ of technifermions in a technicolor (TC) theory give
rise to the masses of the (TC-singlet) quarks and leptons via diagrams
involving exchanges of virtual extended technicolor gauge bosons.

\section{Strongly Coupled QED} 

Our argument is only intended to apply to asymptotically free gauge theories.
However, we offer some remarks on the situation for a particular infrared-free
theory here, namely quantum electrodynamics (QED), based on a U(1) gauge group
with gauge coupling $e$ and some set of fermions $\psi_i$ with charges
$q_i$. Here there are several important differences with respect to an
asymptotically free non-abelian gauge theory.  First, while the chiral limit of
QCD, i.e., quarks with zero current-quark masses, is well-defined because of
quark confinement, a U(1) theory with massless charged particles is unstable,
owing to the well-known fact that these would give rise to a divergent
Bethe-Heitler pair production cross section.  It is therefore necessary to
break the chiral symmetry explicitly with bare fermion mass terms $m_i$.  If
the running coupling $\alpha_1=e^2/(4\pi)$ at a given energy scale $\mu$ were
sufficiently large, $\alpha_1(\mu) \gsim O(1)$, an approximate solution to the
Dyson-Schwinger equation for the propagator of a fermion $\psi_i$ with $m_i <<
\mu$ would suggest that this fermion gains a nonzero dynamical mass $\Sigma_i$
\cite{sd,jbw} and hence, presumably, there would be an associated condensate
$\langle \bar \psi_i \psi_i\rangle$ (no sum on $i$).  However, in analyzing
S$\chi$SB, it is important to minimize the effects of explicit chiral symmetry
breaking due to the bare masses $m_i$. The infrared-free nature of this theory
means that for any given value of $\alpha_1$ at a scale $\mu$, as one decreases
$m_i/\mu$ to reduce explicit breaking of chiral symmetry, $\alpha_1(m_i)$ also
decreases, approaching zero as $m_i/\mu \to 0$.  Since $\alpha_1(m_i)$ should
be the relevant coupling to use in the Dyson-Schwinger equation, it may in fact
be impossible to realize a situation in this theory in which one has small
explicit breaking of chiral symmetry and a large enough value of
$\alpha_1(m_i)$ to induce spontaneous chiral symmetry breaking.  A full
analysis would require knowledge of the bound state spectrum of the
hypothetical strongly coupled U(1) theory, but this spectrum is not reliably
known.

\section{Other Field Theories}

One could also consider supersymmetric SU($N_c$) gauge with a sufficiently
small number $N_f$ of light chiral superfields that the theory is in the phase
with confinement and spontaneous chiral symmetry breaking.  This theory
produces both a gluino and a quark condensate.  With the supersymmetry
unbroken, the physical states are SU($N_c$)-singlet hadrons with degenerate
superpartners.  The extension of our observation for QCD to this theory would
lead one to infer that the various condensates reside within these color
singlet states.

  We also comment on gauge theories in $d=2$ spacetime dimensions. These
include the Schwinger model, QED$_2$ with a massless charged fermion
\cite{schw} and generalizations thereof to QED$_2$ with a set of $N_i$ copies
of massless fermions of different charges $q_i$, and the 't Hooft model
\cite{thooft2d}, namely the limit $N_c \to \infty$ limit, with $g^2N_c$ fixed,
of the ${\rm U}(N_c)_2$ gauge theory with one or more massive or massless
fermions.  These theories have the appeal that they are exactly solvable.
There are well-known differences between them and real QCD, since in $d=2$
dimensions (i) there are no dynamical gauge degrees of freedom, (ii) they are
super-renormalizable, and (iii) the Mermin-Wagner-Coleman (MWC) theorem
\cite{mwc} forbids the breaking of a continuous global symmetry. Nevertheless,
the property that the potential energy associated with a fermion-antifermion
pair separated by a distance $x$ increases linearly with $x$ in a U(1)$_2$ or
${\rm U}(N_c)_2$ gauge theory, and the related absence of any gauge-nonsinglet
states in the spectrum, are reminiscent of confinement in QCD.  Thus, both of
these theories were used as early models exhibiting free behavior at short
distances together with the absence of gauge-nonsinglet physical states, as in
QCD \cite{cks,thooft2d}.  

  The Schwinger model has a spectrum consisting of a free scalar with mass
given by $m^2=e^2/\pi$ and (assuming a particular normal-ordering prescription)
exhibits a nonzero $\langle \bar\psi\psi \rangle \propto m$, and this is a
constant in spacetime. This does not contradict our observation about QCD,
however, because the existence of this $\langle \bar\psi\psi \rangle \ne 0$
depends crucially on the fact that the U(1)$_A$ chiral symmetry is anomalous;
\beq
\partial_\mu J^{\mu 5} = \frac{e^2}{2\pi} \, \epsilon_{\mu\nu}F^{\mu\nu}
\label{u1d2anom}
\eeq
(since otherwise $\langle \bar\psi\psi \rangle$ is forced to vanish by the MWC
theorem).  As is evident from eq. (\ref{u1d2anom}), the necessary condition for
the U(1)$_A$ symmetry to be anomalous is that $F^{01} = E \ne 0$. Recalling
that there are no dynamical gauge fields in $d=2$, one sees that $E$ is a
constant, external electric field.  Hence, in this model $\langle
\bar\psi\psi\rangle$ arises not as a consequence of any dynamical gauge
interactions, but instead as a consequence of the imposition of a nonzero
external electric field, analogous to the Stark effect.  Moreover, the
generalization with $N_i$ copies of massless fermions $\psi_i$ with different
charges $q_i$ is also exactly solvable, and, in agreement with the MWC theorem,
$\langle \bar\psi_i\psi_i\rangle = 0$ (no sum on $i$) if $N_i \ge 2$ (e.g.,
\cite{qed2}).  

  Similar comments apply for the 't Hooft model (with massless fermions);
again, a nonzero condensate is only allowed in the case of one flavor, and it
owes its existence to the fact that the abelian U(1) factor in the ${\rm
U}(N_c) = {\rm SU}(N_c) \times {\rm U}(1)$ gauge group gives rise to an
equation analogous to eq.  (\ref{u1d2anom}), so that, the anomaly in the
axial-vector current is due to a nonzero value of an external chromoelectric
field rather than to any intrinsic dynamics of the theory.

   The existence of condensates which are spacetime constants in models without
confinement, such as the (4D) NJL model and the ($N \to \infty$ limit of the)
2D model with a four-fermion interaction \cite{gn} are fully in agreement with
our observation, since neither of these models exhibits confinement.  

\section{QCD at Finite Temperature} 

So far, we have discussed QCD and other theories at zero temperature.  For QCD
in thermal equilibrium at a finite temperature $T$, as $T$ increases above the
deconfinement temperature $T_{dec}$, both the hadrons and the associated
condensates eventually disappear, although experiments at CERN and BNL-RHIC
show that the situation for $T \gsim T_{dec}$ is more complicated than a weakly
coupled quark-gluon plasma.  Our picture of the QCD condensates here is
especially close to experiment, since, although finite-temperature QCD makes
use of the formal thermodynamic, infinite-volume limit, actual heavy ion
experiments and the resultant (effective) phase transition from confined to
deconfined quarks and gluons takes place in the finite volume and time interval
provided by colliding heavy ions.

This research was partially supported by grants DE-AC02-76SF00515 (SJB) and
NSF-PHY-06-53342 (RS).  SJB thanks Guy de Teramond for useful discussions. 
Preprint SLAC-PUB-13154, YITP-SB-08-07.


\begin{thebibliography}{99}
\bibitem{tmu}
Here and below, unless otherwise stated, we consider theories at temperature
$T=0$ and chemical potential $\mu=0$; in particular, we consider times long
after the QCD deconfinement-confinement phase transition that occurred at $t
\sim 10^{-5}$ sec in the early universe as $T$ decreased below $T_{dec} \simeq
200$ MeV.
\bibitem{ggdual} 
The fact that QCD and resultant strong interactions experimentally conserve 
$P$ and $T$ shows that $P$- and $T$-non-invariant condensates such as 
$\langle G_{\mu\nu}\tilde G^{\mu\nu}\rangle$, where $\tilde G_{\mu\nu} =
(1/2)\epsilon_{\mu\nu\alpha\beta}G^{\alpha\beta}$, are negligible.
(Explaining this is part of the strong CP problem.)
\end{thebibliography}
\end{document}